\documentclass[aps,twocolumn,notitlepage,pra,superscriptaddress,footinbib]{revtex4-1}
\usepackage{amsmath,amssymb,bm}
\usepackage{graphicx}
\usepackage{epstopdf}
\usepackage{latexsym}
\usepackage[normalem]{ulem} 
\usepackage[usenames,dvipsnames]{color}
\usepackage[pdftex,colorlinks=true,citecolor=black,urlcolor=black,linkcolor=black]{hyperref}
\usepackage{natbib}
\usepackage{braket}
\usepackage{comment}
\usepackage{wrapfig}
\usepackage{braket}

\begin{document}

\newcommand{\Rev}[1]{{\color{blue}{#1}\normalcolor}} 
\newcommand{\Com}[1]{{\color{red}{#1}\normalcolor}} 
\title{Simulating a two component Bose-Hubbard model with imbalanced hopping in a Rydberg tweezer array}
\author{Y. Zhang}
\author{A. Gaddie}
\author{H-V. Do}
\author{G.~W. Biedermann}
\author{R.~J. Lewis-Swan}
\affiliation{Homer L. Dodge Department of Physics and Astronomy, The University of Oklahoma, Norman, OK 73019, USA }
\affiliation{Center for Quantum Research and Technology, The University of Oklahoma, Norman, OK 73019, USA}
\date{\today}

\begin{abstract}    
    Optical tweezer arrays of neutral atoms provide a versatile platform for quantum simulation due to the range of interactions and Hamiltonians that can be realized and explored. We propose to simulate a two-component Bose-Hubbard model with power-law hopping using arrays of multilevel Rydberg atoms featuring resonant dipolar interactions. The diversity of states that can be used to encode the local Hilbert space of the Bose-Hubbard model enables control of the relative hopping rate of each component and even the realization of spin-flip hopping. We use numerical simulations to show how multilevel Rydberg atoms provide an opportunity to explore the diverse non-equilibrium quench dynamics of the model. For example, we demonstrate a separation of the relaxation timescales of effective spin and charge degrees of freedom, and observe regimes of slow relaxation when the effective hopping rates of the two components are vastly different due to dynamical constraints arising from hardcore boson interactions. We discuss prospects for studying these effects in state-of-the-art Rydberg tweezer arrays.
\end{abstract}

\maketitle

\section{Introduction}
Programmable tweezer arrays of Rydberg atoms provide a versatile platform for quantum simulation of many models of quantum magnetism in a range of geometries \cite{morgado2021quantum,kaufman2021quantum,barredo2018synthetic,weimer2010rydberg,morgado2021quantum,labuhn2016tunable,de2019observation} and present exciting prospects for realizing exotic quantum phases of matter \cite{vishwanath2021toric,lukin2021liquid,sachdev2021kagome}. By encoding a qubit in Rydberg and/or ground states of an atom, one can selectively realize $1/r^6$ van der Waals or anisotropic $1/r^3$ dipolar interactions that naturally emulate Ising \cite{Schauss2018ising,Labuhn2016ising,monika2020ising,Bernien2017ising,ahn2021ising,kaufman2023ising}, XY \cite{Bornet2023xy,asier2018xy}, or XXZ \cite{signoles2021glassy} spin models. Recent work has demonstrated further versatility with applied fields that realize local or global driving and energy shifts, which can be exploited to realize Floquet Hamiltonians \cite{geier2021ultracold,weidemuller2022floquet,christakis2023probing,Nishad2023floquet}. These tools for Hamiltonian engineering are combined with technical capabilities such as site-resolved imaging and state-preparation in the tweezer apparatus. 

New opportunities are emerging in these systems by applying the aforementioned levels of control to multiple (i.e., more than two) Rydberg levels within each atom. In general, multilevel atoms present a promising platform to expand the frontiers of quantum simulation beyond, e.g., well-established spin-$1/2$ models, and controllably enrich the complexity of many-body dynamics. 
Recent work involving Rydberg atoms has demonstrated how internal structure can be exploited to, e.g., realize synthetic dimensions \cite{ozawa2019topological,hazzard2023synthetic}. For example, a set of neighbouring Rydberg states can be coupled by independent microwaves to study topological physics \cite{kanungo2022realizing}, non-equilibrium dynamics featuring strong interactions \cite{kaden2023rydbergsynthetic}, or classical analog of quantum models \cite{Cohen2024Classical}. 
In this work, we propose to use resonant dipolar exchange interactions between three Rydberg states with alternating parity to simulate the non-equilibrium dynamics of a two component (i.e., spin-$1/2$) Bose-Hubbard model featuring tunable power-law hopping and hardcore interactions. Single atoms are trapped in a programmable tweezer array, realizing the spatial dimension, while the Rydberg levels encode the local Hilbert space of the model, corresponding to the onsite occupation of each boson component. Our mapping explicitly excludes double occupancy of any site to realize perfect hardcore interactions between the effective bosons, while the natural variation of exchange interactions between different levels allows us to easily investigate different hopping rates for each component or even resonant spin-flip tunneling. 
We note that multilevel Rydberg atoms have also similarly been proposed for quantum simulation of a related Bosonic $t-J$ model in a recent work~\cite{Homeier2023Antiferromagnetic}.

The engineered Bose-Hubbard model presents a versatile playground for studying the role of competing timescales and dynamical constraints in the relaxation of an interacting many-body system \cite{Gadway2011glassy,Garrahan2018dynamicalconstraints,Santos2020polargas}. For example, models featuring a mixture of slow and fast (i.e., heavy and light) hopping components on a lattice are of relevance to investigations of anomalously slow relaxation and quasi-localization without onsite disorder \cite{horstmann2007localization,Grover_2014,Muller2015mbl,Papic2015MBL,Yao2016quasimbl,Sirker2019quasimbl,Steinigeweg2020transport,Folling2022massimbalanced,Knap2022transport}. In this spirit, we investigate the quench dynamics of the engineered Bose-Hubbard model as a function of the relative hopping rates and using a variety of tailored initial states that can be prepared in state-of-the-art Rydberg quantum simulators. When the hopping rates are equal, we observe a delineation of the dynamics into effective spin and charge degrees of freedom that is intrinsically related to the power-law behaviour of the the dipolar exchange interaction. On the other hand, when the tunneling rates are strongly imbalanced we observe a separation of timescales in the dynamics, which includes a regime of extremely slow relaxation, and we identify the different stages of thermalization for each component. 

The remainder of the manuscript is organized as follows. In Sec.~\ref{sec:model}, we discuss the details of the two-component Bose-Hubbard model realized by Rydberg tweezer arrays featuring resonant dipolar exchange interactions. Sections \ref{sec:modelud} and \ref{sec:modelbd} then present example quench dynamics, focusing on the role of hopping range and the amplitude for each component. In Sec.~\ref{sec:experiment}, we discuss relevant experimental considerations for the realization of our theoretical predictions and finally in Sec.~\ref{sec:summary} we summarize our results and discuss future directions.  

\section{Quantum simulation with multilevel Rydberg atoms}\label{sec:model}
\begin{figure}[bt]
\begin{center}
\includegraphics[width=0.8\columnwidth]{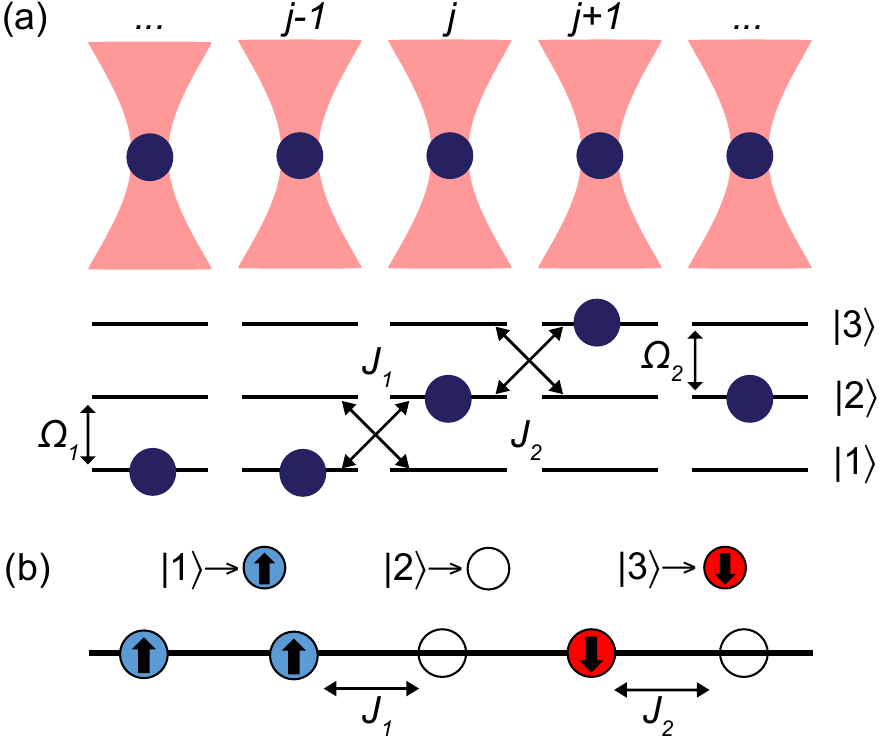}
\caption{(a) Schematic of $1$D optical tweezer array and example level structure. The uniformly spaced tweezers, with position index $j$, each confine a single atom that is prepared in a ladder of three Rydberg levels defined by single-particle states $\vert 1 \rangle$, $\vert 2 \rangle$ and $\vert 3\rangle$. The atoms interact via pairwise dipolar exchange interactions with characteristic strength $J_1$ and $J_2$ between different levels (arrows indicate examples of nearest-neighbour exchange processes -- see text for further details). Microwave fields with Rabi frequencies $\Omega_1$ and $\Omega_2$ coherently couple the single particle states $\ket{1}\leftrightarrow\ket{2}$ and $\ket{2}\leftrightarrow\ket{3}$, respectively, for state preparation and readout. (b) Effective mapping to a model of two-component hardcore bosons. The exchange interaction realizes hopping between sites (indicated by filled and empty circles for occupied and unoccupied sites, respectively) at a rate which depends on the boson component. 
} 
\label{fig:model}
\end{center}
\end{figure}

We consider an array of $\mathcal{N}$ atoms [see Fig.~\ref{fig:model}(a)] individually confined in optical tweezers. Each atom can be prepared in one of a ladder of $n$ Rydberg levels, which we denote by the single-particle states $\vert m \rangle$ where $m = 1,2,...,n$) is the Rydberg state and $j$ indexes the atom (tweezer) in the array. The ladder of Rydberg levels is chosen so that they alternate in parity, i.e., the states $m$ and $m+1$ are separated by a unit increase in orbital angular momentum. We assume that the dominant interactions between the atoms are then resonant dipolar exchange interactions between adjacent Rydberg levels (i.e. states $m$ and $m+1$) \cite{morgado2021quantum}. In addition, a set of resonant microwave fields $\{\Omega_m\}$ can be applied to uniformly couple the single-particle Rydberg states $m$ and $m+1$ across the array [see Fig.~\ref{fig:model}(a)]. Collectively, our system is then described by the general Hamiltonian, 
\begin{align}\label{eq:Hmultilevel}
    \hat{H} &=  \sum_{m=1}^{n-1}\sum_{\substack{i,j=1\\ i\neq j}}^{\mathcal{N}} J^m_{ij} \left(\vert m+1 \rangle \langle m \vert\right)_i\left(\vert m \rangle \langle m+1 \vert\right)_j \nonumber\\ &+ \sum_{m=1}^{n-1}\sum_{i=1}^{\cal N}\frac{\Omega_m}{2}\Big[ \left(\vert m \rangle \langle m+1 \vert\right)_i +\left(\vert m+1 \rangle \langle m \vert\right)_i\Big] ,
\end{align} 
Here, $J^m_{ij} = J_m[1 - 3\cos^2(\theta_{ij})]/r_{ij}^{3}$ characterizes the dipolar interaction between levels $m$ and $m+1$ of atoms in tweezers $i$ and $j$ that are separated by a distance $r_{ij}$ (we assume neighbouring tweezers are separated by unit spacing throughout the manuscript). The angular dependence of the dipolar interaction is set by $\theta_{ij}$, which is the relative angle between the interatomic axis (i.e. the vector $\mathbf{r}_{ij}$ between atoms $i$ and $j$) and the quantization axis set by an externally applied magnetic field. The individual magnitudes $J_m$ of the dipolar interactions are naturally tunable through the specific choice of the Rydberg levels encoding the ladder of single-particle states.

In this work we focus on the case $n=3$, which enables us to use the Rydberg levels to encode the local Hilbert space of a two-component (i.e., spin-$1/2$) Bose-Hubbard model \cite{sundar2019synthetic}, corresponding to the onsite occupation of a single boson of either component or an empty site. The dipolar interactions lead to hopping of the bosons, while limiting to three Rydberg levels engineers a hardcore interaction between the bosons that prevents doubly occupied sites. There are two natural ways to encode these states that lead to a pair of Bose-Hubbard Hamiltonians with features such as different hopping rates for each component and even spin-flip hopping.

The first encoding of the boson occupancies is via the mapping $\ket{1}_j \leftrightarrow \ket{\uparrow}_j$, $\ket{2}_j \leftrightarrow \ket{0}_j$ and $\ket{3}_j \leftrightarrow \ket{\downarrow}_j$, where $\ket{\uparrow}_j$ ($\ket{\downarrow}_j$) corresponds to the site $j$ being occupied with an $\uparrow$ ($\downarrow$) boson and $\ket{0}_j$ encoding an empty site [see Fig.~\ref{fig:model}(b)]. Then, Eq.~(\ref{eq:Hmultilevel}) can be written as a two-component Bose-Hubbard Hamiltonian with component-dependent hopping,
\begin{multline}\label{eq:ud}
    \hat H_{\mathrm{BH}} = \sum_{i\neq j} J^1_{ij}\hat b^\dagger_{\uparrow i} \hat b_{\uparrow j} + J^2_{ij}\hat b^\dagger_{\downarrow i} \hat b_{\downarrow j} \\+\sum_i \frac{\Omega_1}{2}(\hat b^\dagger_{\uparrow i} + \hat b_{\uparrow i}) + \frac{\Omega_2}{2}(\hat b^\dagger_{\downarrow i} + \hat b_{\downarrow i}). 
\end{multline}
Here, $\hat b_{\sigma j}^\dagger$($\hat b_{\sigma j}$) is the creation (annihilation) operator of a spin $\sigma$ boson at site $j$, with $\sigma=\uparrow$ or $\downarrow$. The hardcore interaction enters through the requirement that $\hat b_{\sigma j}^\dagger\hat b_{\sigma' j}^\dagger=\hat b_{\sigma j}\hat b_{\sigma' j}=0$. The first line of Eq.~(\ref{eq:ud}) describes hopping of each component between sites $i$ and $j$ with rates $J^1_{ij}$ and $J^2_{ij}$, respectively, which can be controlled by the choice of Rydberg levels used in the simulator. In the absence of the microwave terms in the second line, which lead to the coherent creation/destruction of bosons at each site, the Hamiltonian separately conserves the total number of $\uparrow$ and $\downarrow$ bosons on the lattice, $N_{\uparrow} = \sum_i \hat b^\dagger_{\uparrow i} \hat b_{\uparrow i}$ and $N_{\downarrow} = \sum_i \hat b^\dagger_{\downarrow i} \hat b_{\downarrow i}$, respectively. This can be understood as a direct result of the conservation of total population in each Rydberg state under the dipolar exchange interaction. 
Note that in the recent work of Ref.~\cite{Homeier2023Antiferromagnetic}, it was proposed to similarly use three Rydberg levels to emulate a bosonic $t-J$ model with hardcore constraints. Our Hamiltonian (\ref{eq:ud}) with $J_{ij}^1=J_{ij}^2$ and $\Omega_1=\Omega_2=0$ corresponds to a limiting case of that work.

Alternatively, encoding the boson occupancies using superpositions of Rydberg levels,  $\ket{\uparrow} \leftrightarrow (\ket{1}+\ket{3})/\sqrt{2}$, $\ket{0} \leftrightarrow \ket{2}$ and $\ket{\downarrow} \leftrightarrow (\ket{1}-\ket{3})/\sqrt{2}$, leads to a distinct Bose-Hubbard Hamiltonian, 
\begin{multline}\label{eq:bd}
    \hat H^{\prime}_{\mathrm{BH}} = \sum_{i\neq j} J^h_{ij} \left( \hat b^\dagger_{\uparrow i} \hat b_{\uparrow j} + \hat b^\dagger_{\downarrow i} \hat b_{\downarrow j} \right) + 
    J^f_{ij} \left( \hat b^\dagger_{\uparrow i} \hat b_{\downarrow j} + \hat b^\dagger_{\downarrow i} \hat b_{\uparrow j} \right) \\+\sum_i \frac{\Omega_{\uparrow}}{2}(\hat b^\dagger_{\uparrow i} + \hat b_{\uparrow i}) + \frac{\Omega_{\downarrow}}{2}(\hat b^\dagger_{\downarrow i} + \hat b_{\downarrow i}) .
\end{multline}
where $J^h_{ij} = (J^1_{ij}+J^2_{ij})/2$, $J^f_{ij} = (J^1_{ij} - J^2_{ij})/2$, $\Omega_{\uparrow} = (\Omega_1 + \Omega_2)/\sqrt{2}$ and $\Omega_{\downarrow} = (\Omega_1 - \Omega_2)/\sqrt{2}$. The first line of the Hamiltonian (\ref{eq:bd}) describes a pair of composite hopping processes between bosons at sites $i$ and $j$: i) hopping of each component identically at a rate $J^h_{ij}$ and ii) spin-flip hopping at a rate $J^f_{ij}$ that converts between each component. The latter can be viewed as a form of spin-orbit coupling \cite{Syzranov2014spinorbit,lienhard2020spinorbit}, where the spin of a hardcore boson repeatedly flips as it propagates spatially through the array. In the absence of the microwave driving, the Hamiltonian (\ref{eq:bd}) features different conserved quantities depending on the values of $J_1$ and $J_2$: If they are equal, so that $J^f_{ij} = 0$, then the total number of $\uparrow$ and $\downarrow$ bosons are separately conserved, but if the interactions are different, so that $J^f_{ij} \neq 0$ and spin-flips accompany the hopping, then only the total boson number is conserved.

In the following sections, we numerically investigate the quench dynamics of each model, with a particular focus on the relaxation dynamics at long times. We show that dynamical constraints on the hopping dynamics in $1$D due to the hardcore interactions, in combination with symmetries and associated conserved quantities of each model, present key ingredient for slow and multi-step relaxation that can be characterized with both local observables and the growth of entanglement.

\section{Quench dynamics of $\hat{H}_{\mathrm{BH}}$}\label{sec:modelud}
We first study the quench dynamics of the Hamiltonian $\hat{H}_{\mathrm{BH}}$ [Eq.~(\ref{eq:ud})]. We limit our investigation to $1$D and assume $\Omega_1 = \Omega_2 = 0$. This enables both a clear understanding of the constraints on the system dynamics due to, e.g., the hardcore interactions, but also allows us to exploit the conservation of the total boson number for each component to reduce the dimension of the Hilbert space and compute exact dynamics for system sizes $\mathcal{N} \leq 16$. For generality, we consider the hopping term in Eq.~(\ref{eq:ud}) to be described by a power-law $J^m_{ij} = J_m/r_{ij}^{\alpha}$ (note that the angular dependence of the dipolar interaction is absorbed by the hopping amplitude $J_m$ in $1$D), where $\alpha\in[0,\infty)$ controls the hopping range.

In the following, we show that the relaxation dynamics can be split into two characteristic regimes. First, when the hopping rates of each component are equal, the dynamics is strongly influenced by the power-law dependence of the hopping. We give an explanation for this by dividing the system into spin and charge degrees of freedom, which feature distinct relaxation timescales that depend on the power-law exponent of the hopping. Second, when the hopping rates are strongly disparate we identify a regime of slow relaxation due to competition between the transport of each component through the array. We investigate each of these regimes separately and use specifically designed initial states to amplify the relevant behaviour.

\subsection{Power-law hopping and spin-charge dynamics}\label{sec:longrange}
While a natural basis for the two-component boson system is given by the occupancies of each component at a given site, one can equivalently describe the system using effective spin and charge (density) degrees of freedom. Specifically, any state in the accessible Hilbert space can be uniquely labelled by the spatial ordering of the two components over the occupied sites (i.e., spin ordering), combined with the locations of the occupied sites (i.e., charge). For example, a state given in the original two-component occupancy basis as $\ket{\psi} = \ket{\uparrow \downarrow 0}$, corresponding to an $\uparrow$ boson in site $1$, $\downarrow$ boson in site $2$ and an empty site $3$, could be equivalently written as $\ket{\psi} \equiv \ket{\uparrow \downarrow}_{s} \otimes \ket{1 2}_{c}$ where the first ket encodes the spin ordering and the second the charge configuration. 

In the case where hopping is restricted to nearest-neighbour sites $i$ and $j$ in a $1$D lattice, the spin-charge description becomes particularly powerful: The spin ordering is a conserved quantity, as the hardcore bosons cannot hop past each other, and the dynamics is entirely described by the charge dynamics driven by nearest-neighbour hopping. In the case where the hopping rates are equal, $J_1 = J_2$, this conservation of spin ordering can be exploited to provide an exact solution for the model \cite{fuchs20051dspinor,zhang2018impenetrable,zhang2019quantum}. However, power-law hopping, such as that featured in Eq.~(\ref{eq:ud}) due to the underlying dipolar interaction between Rydberg atoms, introduces next-nearest-neighbour hopping and leads to a breakdown of the spin ordering in the dynamics. Nevertheless, when the power-law hopping is sufficiently short-range, the fundamentally different dependence of the charge and spin degrees of freedom on the nearest- and next-nearest-neighbour hopping rates leads to a separation of the dynamics and associated timescales for spin and charge. As a comment, we note that the fact that the power-law dipolar interaction of the underlying multilevel Rydberg atom system is crucial to the non-equilibrium dynamics is in contrast to prior discussions of ground-state physics that fruitfully approximated the interaction as nearest-neighbour \cite{sundar2019synthetic}.

To study the delineation into spin and charge dynamics, we initiate a quench from an initial charge density wave with superimposed Neel ordering of the spin components, $\ket{\varphi_{\mathrm{cdw}}} = \ket{\uparrow\downarrow0\uparrow\downarrow0\cdots}$. Furthermore, to better understand the role of interaction range in the spin-charge picture, we consider power-law hopping with $\alpha \in [0,6]$ and equal amplitudes $J_1=J_2=J$. The strength of nearest- and next-nearest-neighbor hopping is distinguished by a factor of $1/2^\alpha$. To minimize trivial effects from the edges of the $1$D chain we use periodic boundary conditions~\footnote{Operationally, this is achieved by using a ring geometry and artificially defining $r_{ij}$ as the arc length between site $i$ and $j$ to mimic $1$D. The former aspect is motivated by a recent experimental realization of periodic boundaries with tweezers \cite{weidemuller2022floquet}.}. All results in the following are obtained by numerically integrating the Schr\"{o}dinger equation with the Krylov subspace method \cite{IterativeMethodBook2023}.

\begin{figure}[bt]
\begin{center}
\includegraphics[width=1.0\columnwidth]{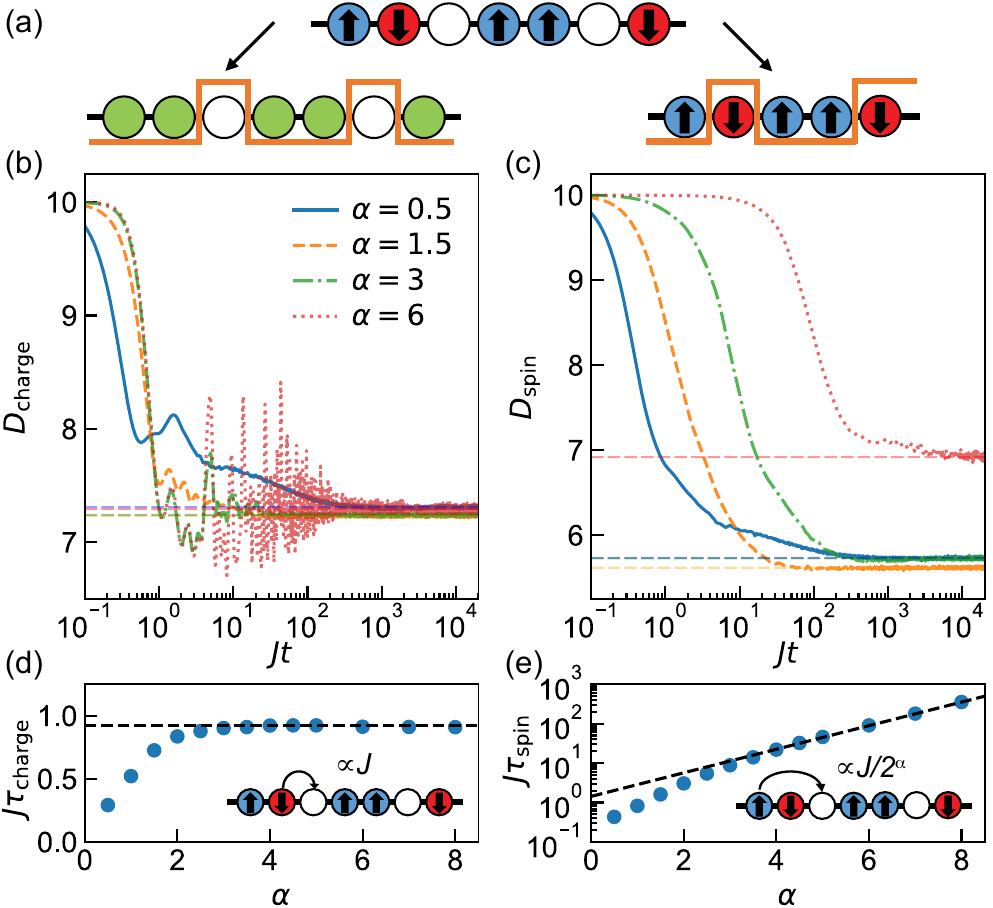}
\caption{Quench dynamics for initial charge density wave state $\ket{\varphi_{\mathrm{cdw}}} = \ket{\uparrow\downarrow0\uparrow\downarrow0\cdots}$ with periodic boundary conditions and ${\cal N}=15$. (a) Illustration of the definition of the number of charge domains $D_{\rm charge}$ [Eq.~(\ref{eq:dd}), left] and spin domains $D_{\rm spin}$ [Eq.~(\ref{eq:sd}), right]. 
(b) and (c) Evolution of $D_{\rm charge}$ and $D_{\rm spin}$ as a function of $Jt$. The plotted results indicate different hopping exponents $\alpha$ [see legend in (b)]. Horizontal dashed lines in both panels indicate the long-time average value of $D_{\rm charge, f}$ and $D_{\rm spin, f}$ at $Jt > 10^4$ (see main text). (d) and (e) Characteristic time $\tau_{\rm charge}$ and $\tau_{\rm spin}$ (see main text) extracted from data shown in (c) and (d), respectively. The dashed line in (d) indicates $\tau_{\rm charge}=0.92$, which is the value for $\alpha=4$, to guide the eye. The dashed line in (e) is $\tau_{\rm spin} = (1.99)^\alpha$, which is obtained via a power-law fit to the $\alpha\geq 4$ data. Insets in panels (d) and (e) indicate the nearest-neighbour and next-nearest neighhbouring hopping processes that are expected to dominate the charge and spin timescales, respectively, for large $\alpha$.
} 
\label{fig:longrange}
\end{center}
\end{figure}

The dynamics of the spin ordering can be captured by the observable, 
\begin{equation}\label{eq:sd}
    \hat D_{\rm spin}=\frac{1}{2}\sum_{j\in{squeezed}}(1-\hat S^z_j\hat S^z_{j+1})\, .
\end{equation}
where the summation $j$ runs through the `squeezed space' \cite{hilker_2017_squeezed,bohrdt_2021_squeezed} of only sites that are occupied by a boson, and $\hat S^z_j|\uparrow\rangle_j=|\uparrow\rangle_j$ ($\hat S^z_j|\downarrow\rangle_j=-|\downarrow\rangle_j$). This quantity corresponds to counting the number of spin domain walls (e.g., boundaries between $\uparrow$ and $\downarrow$ spins) after removing the empty sites of the $1$D chain [see Fig.~\ref{fig:longrange}(a), right]. When spin ordering is conserved the number of spin domain walls will remain constant. 
To track the charge dynamics we similarly define
\begin{equation}\label{eq:dd}
    \hat D_{\rm charge}=\frac{1}{2}\sum_{j}[1-(2\hat n_j-1) (2\hat n_{j+1}-1)]\,,
\end{equation}
where $\hat n_j=\hat b^\dagger_{\uparrow j} \hat b_{\uparrow j}+\hat b^\dagger_{\downarrow j} \hat b_{\downarrow j}$. This quantity tracks the number of charge domain walls, which are defined as the boundary between an empty and occupied site, regardless of the spin state of the latter [see Fig.~\ref{fig:longrange}(a), left].

In Figs.~\ref{fig:longrange}(b) and (c), we plot the evolution of $D_{\rm charge}$ and $D_{\rm spin}$ for $\alpha=0.5$ (solid lines), $1.5$ (dashed lines), $3$ (dashed dotted lines), and $6$ (dotted lines). For the initial state $\ket{\varphi_{\mathrm{cdw}}}$, we identically have $D_{\rm spin}=D_{\rm charge}=2\mathcal{N}/3$ at $t=0$. 
Just after the quench, we generically observe a decrease of $D_{\rm charge}$ away from its initial value over timescales $Jt \sim 1$. Notably, the initial dynamics for $\alpha = 3$ and $\alpha = 6$ are virtually indistinguishable, which is consistent with the expectation that the charge dynamics is dominated by nearest-neighbour hopping. The decay of the charge domains is slightly accelerated for $\alpha = 0.5$ and $1.5$, which we attribute to corrections from long-range hopping processes. At intermediate times $Jt \gtrsim 1$, $D_{\rm charge}$ shows distinct behaviour depending on the hopping exponent $\alpha$. For instance, when considering larger values (e.g., $\alpha = 6$), we often observe significant oscillations of $D_{\rm charge}$ around its long-term average. This can be understood by noting that the model is integrable in the extreme case of $\alpha \to \infty$. For $\alpha = 0.5$ we also observe a regime of comparatively slow relaxation over multiple decades in time. In all cases, the duration of this intermediate regime appears to be correlated with the relaxation of the spin degree of freedom (see below), demonstrating that the detailed relaxation of the charge degree of freedom cannot be completely disentangled from the spin dynamics. At long times, $D_{\rm charge}$ relaxes to a nearly common value regardless of $\alpha$ that is consistent with expectations based on a diagonal ensemble calculation (see Appendix \ref{sec:app1}). This value is also close to the one predicted from an infinite temperature canonical ensemble, $D_{\rm charge} \to (N_\uparrow+N_\downarrow)^2/(\mathcal{N}-1) \simeq 7.14$, to which expectation values of local observables should thermalize at long times in the limit of large system size. 

On the other hand, the decay of $D_{\rm spin}$ in Fig.~\ref{fig:longrange}(c) shows an initial delay that grows with $\alpha$, which is consistent with the importance of beyond nearest-neighbour hopping for changes to the spin ordering. To quantify this we investigate the relaxation timescales $\tau_{\rm spin}$ and $\tau_{\rm charge}$ for the spin and charge degrees of freedom, which are defined as the time over which the number of domain walls has relaxed halfway to its long time limit. For the spin degree of freedom this is given by $D_{\rm spin}(\tau_{\mathrm{spin}}) = \left[D_{\rm spin}(0) + D_{\rm spin, f}\right]/2$. Ideally, $D_{\rm spin, f}$ is obtained by an exact computation of the diagonal ensemble, but for our system size it is more efficient to estimate this value using a long-time average $D_{\rm spin, f} = \frac{1}{T}\int_{T}^{2T} D_{\rm spin}(t) dt$ with $T = 10^4 J^{-1}$. The relaxation of the number of charge domain walls is analyzed identically. 

We plot $\tau_{\rm charge}$ and $\tau_{\rm spin}$ as a function of the interaction exponent $\alpha$ in Figs.~\ref{fig:longrange}(d) and (e). For both timescales, we observe a delineation into approximately short- and long-range regimes, characterized by the magnitude of the hopping exponent $\alpha$. For $\alpha \gtrsim 2$ the dynamics is dominated by the shortest-range hopping process relevant for the specific degree of freedom, evidenced by the fact that spin relaxation timescale grows exponentially as $J\tau_{\rm spin} \approx 2^{\alpha}$ while the charge relaxation is characterized by a constant $J\tau_{\rm charge} \approx 0.92$. The former result demonstrates that the breakdown of spin-ordering in the initial state $\ket{\varphi_{\mathrm{cdw}}}$ is dominated by the $\alpha$-dependent next-nearest neighbour hopping rate $J/2^{\alpha}$, while the charge dynamics occurs at a fixed nearest-neighbour rate $J$. On the other hand, for $\alpha \lesssim 2$ the hopping becomes long-range and the delineation of the spin and charge dynamics vanishes with $\tau_{\mathrm{spin}}/\tau_{\mathrm{charge}} \approx 1$ as $\alpha \to 0$.

Interestingly, the disparate relaxation of the spin and charge is also observable in the growth of bipartite entanglement after the quench. We quantify this using the von Neumann entanglement entropy, 
\begin{equation}\label{eq:entanglement}
     S=-\mathrm{Tr} \rho_A\ln(\rho_A)\, ,
\end{equation}
where $\rho_A=Tr_{\bar A}\rho$ ($\rho=|\varphi(t)\rangle\langle\varphi(t)|$) is the reduced density matrix obtained by carrying out a partial trace of the system split into a bipartition of $A$ and $\bar A$ that we construct in different ways. One is a spatial bipartition, with associated entanglement entropy $S_{\rm vN}$, where the Hubbard lattice is separated into equal pieces such that $A$ includes sites $1$ to $\lfloor {\cal N}/2\rfloor$ and $\bar{A}$ contains the remainder. Alternatively, we consider a bipartition in terms of the spin and charge degrees of freedom \cite{Zheng2018spincharge}, which we denote by the spin-charge entanglement entropy $S_{\rm sc}$. While $S_{\rm vN}$ captures the spread of entanglement across the lattice, $S_{\rm sc}$ instead captures the build up of entanglement between the charge and spin degrees of freedom across the whole chain.

\begin{figure}[bt]
\begin{center}
\includegraphics[width=1\columnwidth]{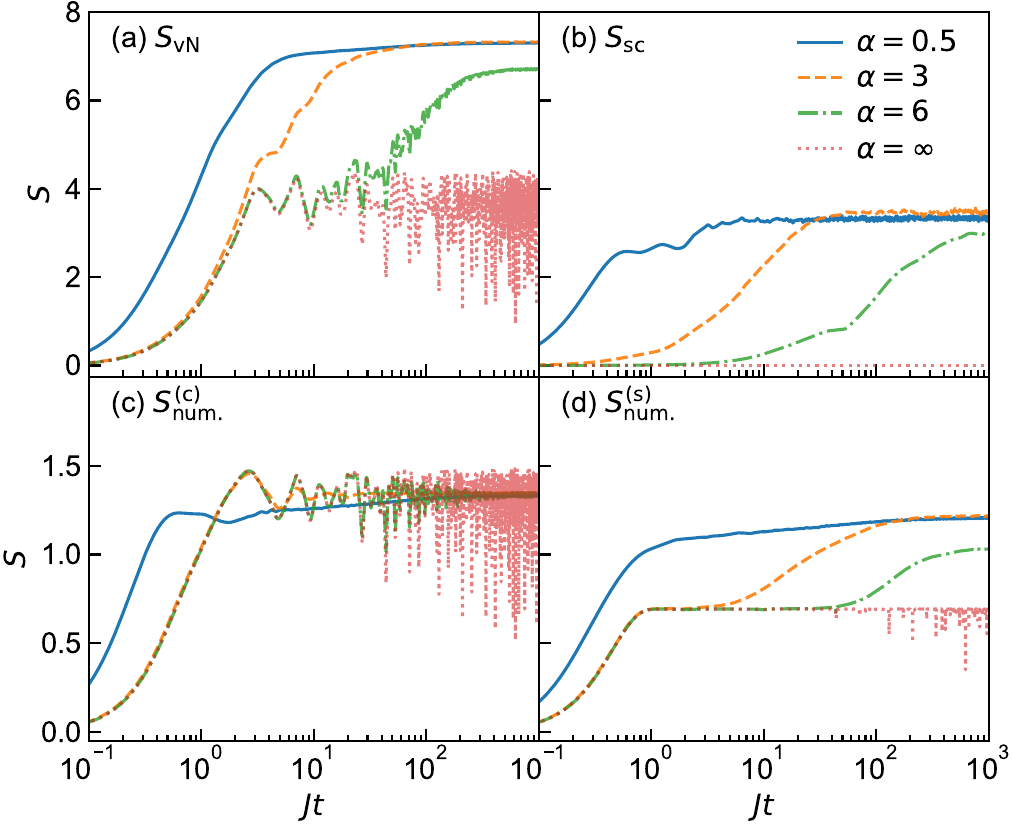}
\caption{Growth of entanglement starting from initial charge density wave state $\ket{\varphi_{\mathrm{cdw}}} = \ket{\uparrow\downarrow0\uparrow\downarrow0\cdots}$ with periodic boundary conditions and ${\cal N}=15$. We plot (a) spatial entanglement entropy $S_{\rm vN}$, (b) spin-charge entropy $S_{\rm sc}$, (c) charge number entropy $S_{\rm num.}^{\rm (c)}$ and (d) spin number $S_{\rm num.}^{\rm (s)}$. See main text for definitions and choice of bipartitions. All panels show data for hopping exponents $\alpha=0.5$ (solid line), $3$ (dashed line), $6$ (dashed dotted line) as well as $\alpha=\infty$ (corresponding to nearest-neighbor hopping, dotted line).} 
\label{fig:Ssc}
\end{center}
\end{figure}

In Fig.~\ref{fig:Ssc}(a), we plot $S_{\rm vN}$ for $\alpha=0.5$ (solid line), $3$ (dashed line) and $6$ (dashed dotted line), respectively, along with nearest-neighbour hopping ($\alpha = \infty$, dotted line) as a reference case. We observe a separation of the entanglement dynamics into two timescales for sufficiently short-range hopping. At short times $Jt \lesssim 1$, the entanglement entropy grows nearly identically for the cases with $\alpha \geq 3$, reflecting that over these timescales the growth of entropy is dominated by the decay of the initial charge density wave throughout the chain [recall Fig.~\ref{fig:longrange}]. The subsequent behaviour of $S_{\rm vN}$ depends on the specific value of $\alpha$ and the effective dimension of the Hilbert space that the system is able to access. For example, for nearest-neighbour hopping the entanglement has effectively saturated, albeit with large fluctuations at late times due to the integrability of the model. For $\alpha = 6$ we observe a similar transient period where $S_{\rm vN}$ weakly oscillates around a fixed value identical to $\alpha = \infty$, but a second period of growth starts at $Jt\gtrsim50$ ($\sim 2^6$) after which the entanglement finally saturates. This second stage is associated with the breakdown of the Neel order of the spins: Initially, the quasi-saturation of the entanglement is due to system being constrained to a small sector of the Hilbert space defined by the fixed spin-order, but at late times the entanglement grows again as the spin-ordering is broken by next-nearest neighbour hopping and the system probes the full Hilbert space. Signatures of the separation of spin and charge timescales can also still be seen in the dipolar case, $\alpha = 3$, although they are less striking. On the other hand, for long-range hopping, $\alpha = 0.5$, the entanglement grows to near the maximum value without any apparent distinction between spin or charge regimes, as the system is able to access the full Hilbert space even at short times.

Further support for this understanding of the spatial entanglement growth is shown in Fig.~\ref{fig:Ssc}(b), wherein we plot the corresponding results for the spin-charge entanglement $S_{\rm sc}$ (Eq.~\ref{eq:entanglement}). For short-range hopping ($\alpha=3, 6$ and $\infty$) $S_{\rm sc}$ remains zero for a transient period, before an increase that commences in tandem with the second rise of $S_{\rm vN}$ seen in panel (a). This is consistent with the fact that at short times, when the spin-ordering is fixed, there can be no appreciable growth of the spin-charge entanglement as the wavefunction factorizes in the spin and charge basis. Conversely, this cannot be true when the spin and charge dynamics are mixed together, such as the case for long-range hopping $\alpha=0.5$.

In general, obtaining the entanglement entropy is a challenging task in any quantum simulation platform. For this reason, we also study the behaviour of the closely related number entropy associated with the charge and spin. We define the charge number entropy as \cite{lukinMBL2019,melko2016entang,wiseman2003entang}, 
\begin{equation}\label{eq:Snumber}
     S^{\rm (c)}_{\rm num.} =-\sum_{n}p^{(A)}_n\ln p^{(A)}_n,
\end{equation}
where $p^{(A)}_n$ is the probability of finding a total of $n=n_\uparrow+n_\downarrow$ atoms in subregion $A$ of the lattice. Similarly, the spin number entropy is given by, 
\begin{equation}\label{eq:Snumber2}
     S^{\rm (s)}_{\rm num.} =-\sum_{n_\uparrow}p^{(\cal A)}_{n_\uparrow}\ln p^{(\cal A)}_{n_\uparrow} . 
\end{equation}
Here, $p^{(\cal A)}_{n_\uparrow}$ is the probability of finding $n_\uparrow$ bosons in a subregion $\cal A$, where we define a bipartition of the system in squeezed space into regions $\cal A$ and $\bar{\cal A}$. We always choose this bipartition so that $\cal A$ encompasses sites $1$ to $\lfloor (N_\uparrow+N_\downarrow)/2\rfloor$~\footnote{Note that there is an ambiguity of labelling lattice sites in squeezed space due to the periodic boundary conditions. We choose to always specify site $1$ relative to a fixed location in the full spatial lattice.}. 

We show the results for $S^{\rm (c)}_{\rm num.}$ and $S^{\rm (s)}_{\rm num.}$ in Figs.~\ref{fig:Ssc}(c) and (d). The charge number entropy grows identically for short-range hopping, $\alpha =3,6,\infty$ before saturating towards a nearly common value. This is consistent with the interpretation of the first rise of $S_{\mathrm{vN}}$ being dominated by charge dynamics. The nearest-neighbour data shows large oscillations at longer times due to the integrability of the model in that case. 
The dynamics of $S^{\rm (s)}_{\rm num.}$ requires a more careful discussion. The short time growth of $S^{\rm (s)}_{\rm num.}$ for $Jt \lesssim 1$ is an artifact of the ambiguity of defining a bipartition in the squeezed space, as a result of the periodic boundary conditions and the discrete translational symmetry of our initial state. However, the growth at longer times is a direct consequence of the breakdown in spin ordering, and sets in at approximately the same time as the period of second growth in the full calculation of $S_{\rm vN}$.

\subsection{Effects of imbalanced hopping}\label{sec:hopping}
In Rydberg atoms, there are a plethora of states that can be used to encode the boson occupancies, and thus an intrinsically large variation in the associated dipolar interactions. Motivated by this, in this section we turn our attention to quench dynamics featuring distinct hopping amplitudes $J_1\neq J_2$. We find that the delineation of the dynamics into charge- and spin-dominated regimes quickly vanishes upon breaking the symmetry between the components, and in the extreme limit $J_1 \ll J_2$ leads to a new regime featuring a dramatic slowdown in relaxation. The latter effect arises due to a combination of the imbalanced hopping rates, hardcore interactions and power-law hopping range, and overlaps with recent studies of slow relaxation due to dynamical constraints \cite{Garrahan2018dynamicalconstraints,Muller2015mbl,Yao2016quasimbl,Sirker2019quasimbl,Folling2022massimbalanced}. Without loss of generality, we will assume in the following that $J_1/J_2 \leq 1$ and therefore refer to $\ket{\downarrow}$ ($\ket{\uparrow}$) as the fast (slow) hopping component. In contrast to the prior section, we primarily focus our discussion and analysis on the experimentally relevant case of dipolar interactions with $\alpha = 3$.

\begin{figure}[bt]
\begin{center}
\includegraphics[width=1\columnwidth]{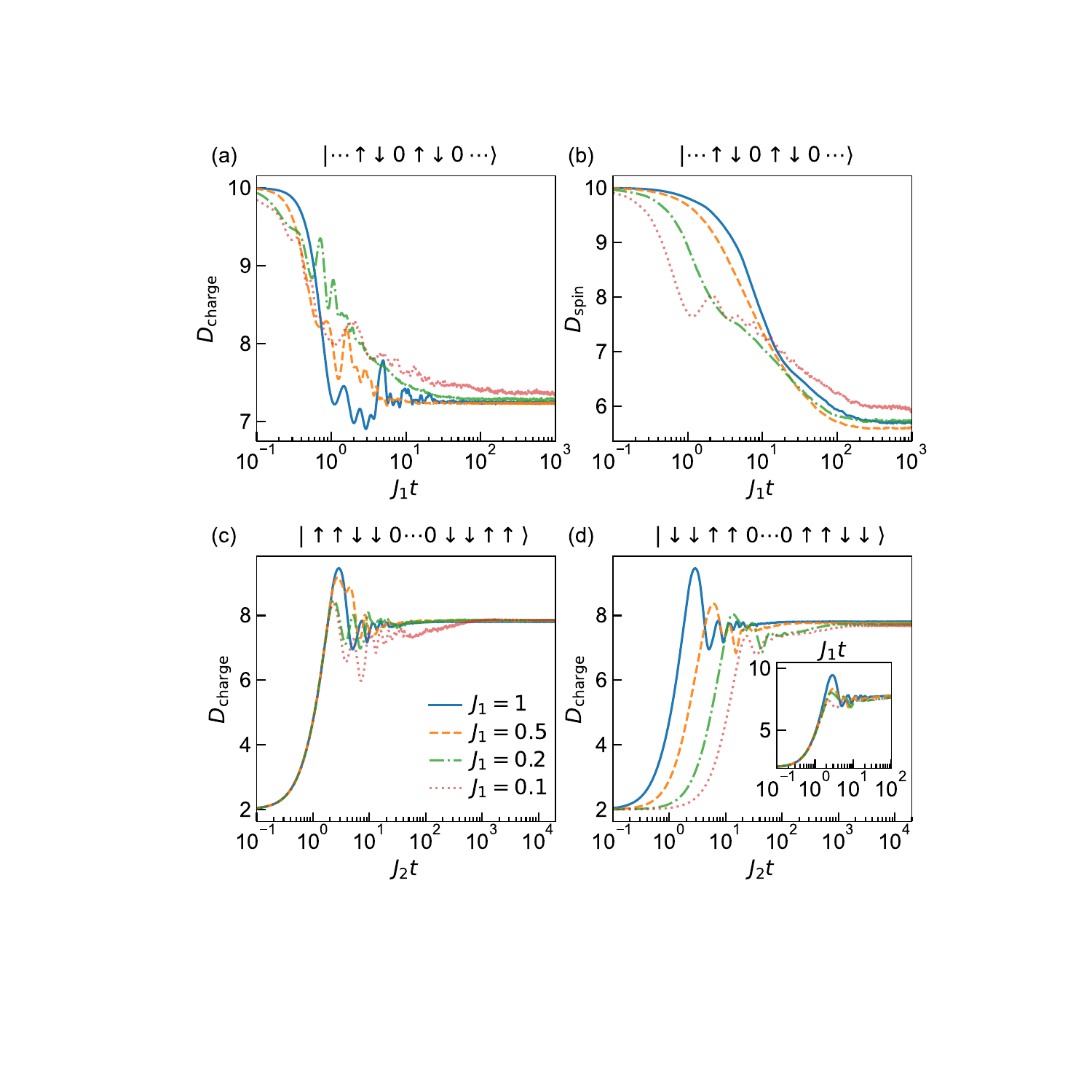}
\caption{Quench dynamics with imbalanced hopping. All panels use fixed $J_2 = 1$ and vary $J_1$ [see legend of panel (c)]. (a) and (b) Relaxation of $D_{\rm charge}$ and $D_{\mathrm{spin}}$ as a function of rescaled time $J_1 t$. Calculations are for an initial charge density wave state $\ket{\varphi_{\rm CDW}}$ (indicated above panels) with periodic boundary conditions and ${\cal N}=15$. (c) and (d) Relaxation of $D_{\rm charge}$ as a function of time $J_2 t$ for different initial charge domain states (indicated above panels) with ${\cal N}=16$ sites and open boundary conditions. In the inset of (d) we plot $D_{\rm charge}$ as a function of rescaled time $J_1 t$.}   
\label{fig:unequalhopping}
\end{center}
\end{figure}

We begin by investigating the dynamic of spin and charge observables for moderately imbalanced hopping amplitudes in the range $0.1 \leq J_1/J_2 < 1$ (we fix the value of $J_2$ and vary $J_1$ throughout). Panels (a) and (b) of Fig.~\ref{fig:unequalhopping} show the dynamics of $D_{\rm charge}$ and $D_{\rm spin}$ starting from the same charge-density wave state $\ket{\varphi_{\mathrm{CDW}}}$. When the hopping amplitudes are comparable (e.g., dashed lines with $J_1/J_2=0.5$), the delineation of the spin and charge dynamics is preserved and qualitatively matches the prior results shown in Fig.~\ref{fig:longrange}. However, as $J_1/J_2$ is decreased, the initial dynamics of $D_{\rm spin}$ are accelerated and the characteristic delay relative to the charge relaxation vanishes. This is understood by noting that when $J_1/J_2 \lesssim 2^{-3}$ the nearest-neighbour hopping rate of the slow component becomes comparable to the next-nearest-neighbour hopping rate of the fast component, and thus there is no longer any fundamental distinction between the spin and charge timescales. Separately, we note that for the smallest $J_1/J_2=0.1$, both $D_{\rm charge}$ and $D_{\rm spin}$ relax noticeably slower than for other $J_1/J_2$ values (note the rescaling of the time axis in both panels). We discuss this regime in more detail below.

Figures \ref{fig:unequalhopping}(c) and (d) additionally show that imbalanced hopping amplitudes naturally lead to a dependence of the charge relaxation dynamics on the initial spin ordering. We illustrate this by using a pair of initial states composed of separated charge domains at the boundaries of the lattice, but with mirroring arrangements of the initial spin order (see panel insets). Depending on whether the slow hopping components initially occupy the exterior [panel (c)] or interior [panel (d)] sites of the domains, the initial dynamics of $D_{\rm charge}$ is set by either $1/J_2$ or $1/J_1$, respectively. For the latter case, the apparent slowdown in the initial growth of $D_{\rm charge}$ is due to blocking of the transport of the fast hopping component due to the hardcore inter-component interaction. In addition, the late-time relaxation of $D_{\rm charge}$ can also show sharply distinct timescales depending on the initial spin ordering [compare, e.g. $J_1=0.2$, dashed dotted lines in (c) and (d)].

\begin{figure}[bt]
\begin{center}
\includegraphics[width=1\columnwidth]{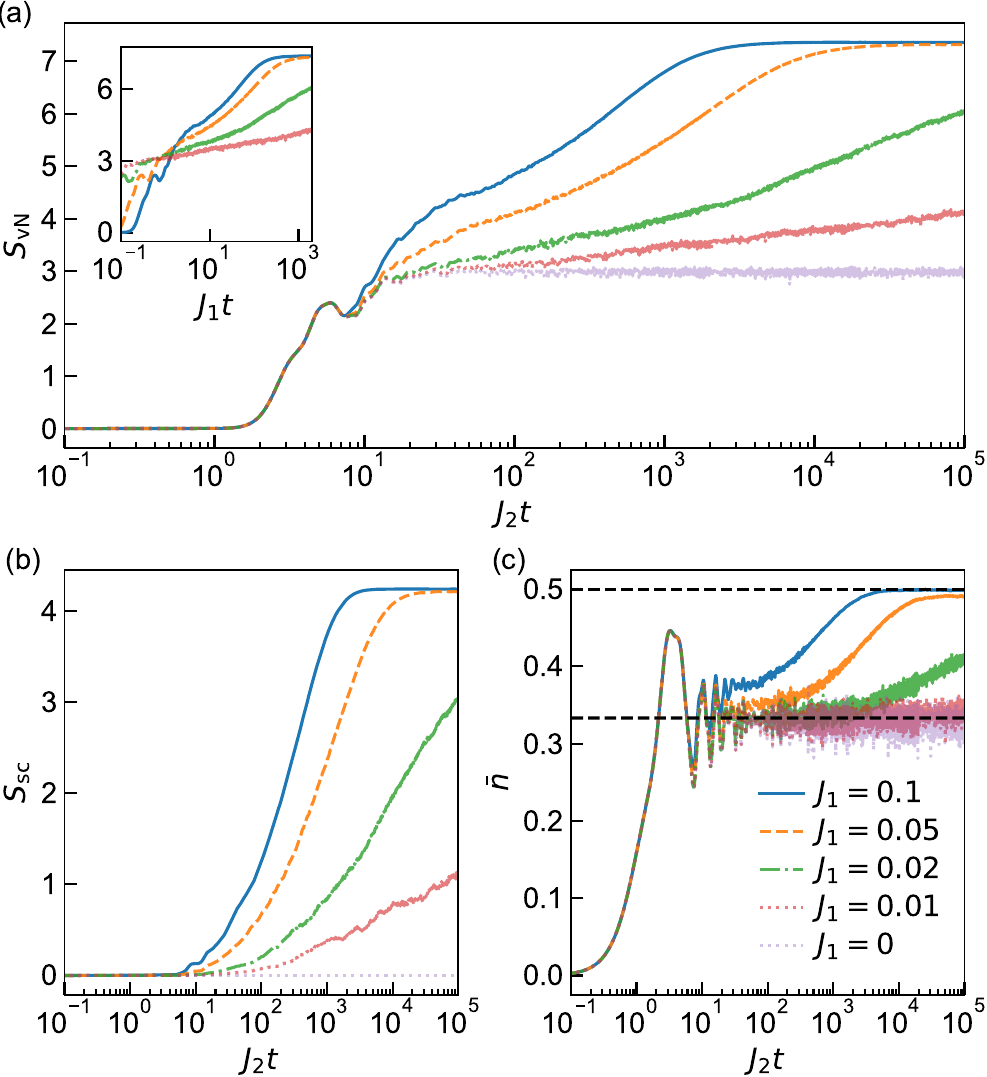}
\caption{Quench dynamics for strongly imbalanced hopping with initial state $\ket{\uparrow\uparrow\downarrow\downarrow0\cdots0\downarrow\downarrow\uparrow\uparrow}$ and $\mathcal{N} = 16$. All panels use fixed $J_2 = 1$ and vary $J_1$ [see legend of panel (c)]. (a) and (b) Evolution of spatial and spin-charge entanglement entropy, $S_{\rm vN}$ and $S_{\rm sc}$, respectively, as a function of time $J_2 t$. 
We also plot $S_{\rm vN}$ as a function of rescaled time $J_1t$ in the inset of panel (a). 
(c) The local charge density $\bar{n} = \frac{1}{(\mathcal{N}/2)}\sum_{j=\mathcal{N}/4}^{3\mathcal{N}/4} \hat{n}_j$ averaged over the initially empty central sites. Horizontal dashed lines indicate $\bar n= N_{\downarrow}/(\mathcal{N}-N_{\uparrow}) = 1/3$ and $\bar n = (N_{\downarrow} + N_{\uparrow})/\mathcal{N} = 1/2$, respectively. The system size is set as ${\cal N}=16$ with open boundary conditions.} 
\label{fig:slowJ1}
\end{center}
\end{figure}

The interplay of the two components in the long-time relaxation dynamics is particularly magnified in the extreme case $J_1/J_2 \ll 1$. Figure \ref{fig:slowJ1} shows quench dynamics for the initial domain state $\ket{\uparrow\uparrow\downarrow\downarrow0\cdots0\downarrow\downarrow\uparrow\uparrow}$ wherein the slow hopping component initially occupies the edges of the lattice. Panel (a) shows the entanglement entropy $S_{\mathrm{vN}}$ of a spatial bipartition of the lattice and features a striking two-step evolution that depends on the ratio $J_1/J_2$. In the extreme limit $J_1/J_2 = 0$ (magenta dotted line), the system reduces to a single component Bose-Hubbard model that exists on a sublattice of $\mathcal{N}-N_{\uparrow}$ sites. Entanglement entropy grows linearly on a timescale $\propto 1/J_2$ and saturates at a nominal value set by the smaller Hilbert space of the single-component system. For small $J_1/J_2 \neq 0$ the initial dynamics remains identical, as the fast component in the interior of the chain quickly equilibrates on the same timescale $\propto 1/J_2$ with the entanglement entropy approaching the single-component value. However, a second stage of dynamics commences at $J_2t \gtrsim 10$ that involves the hopping of the slow component. This second stage is characterized by slow growth of the entanglement entropy over multiple decades in time before eventual saturation to a larger value reflective of the increased dimension of the two-component Hilbert space. The timescales for the second-stage are not set simply by $1/J_1$ [see rescaled data in inset of panel (a)] nor does the data collapse for a more general rescaling of time by $J_1^{\beta} t$ with $\beta$ an arbitrary power (see below). Similar two-step behaviour is observed in the spin-charge entropy [panel (b)] and more readily accessible observables such as the local charge density [panel (c)]. The former remains near zero until $J_2 t \sim 10$, reflecting the initial preservation of the spin-order, before a period of slow growth similar to $S_{\mathrm{vN}}$. On the other hand, the local charge density $\bar{n} = \frac{1}{(\mathcal{N}/2)}\sum_{j=\mathcal{N}/4}^{3\mathcal{N}/4} \hat{n}_j$ averaged over the initially empty central sites, quickly grows to $\bar{n} \approx N_{\downarrow}/(\mathcal{N}-N_{\uparrow})$ -- which is the expected homogeneous density of a single component across the subset of available sites -- on timescales $1/J_2$ before slowly increasing towards $\bar{n} \approx (N_{\downarrow} + N_{\uparrow})/\mathcal{N}$ as the charge density becomes uniform across the lattice.

The second-stage of slow dynamics that we observe arises not simply due to the disparity of the hopping rates, but as a result of a dynamical constraint on the hopping dynamics due to the hardcore interaction, combined with the power-law dependence of the hopping that plays a crucial role in relaxation at long times (which includes rearrangements of the spin-ordering). To understand this, we first point out that if the hopping process is all-to-all ($\alpha = 0$), the geometry of the lattice is irrelevant to the dynamics and the hopping of each component is effectively independent of the spin configuration. This leads to relaxation that depends only on the slow hopping rate $J_1$, which we demonstrate in Appendix \ref{sec:app2} by showing that the entanglement dynamics collapses when rescaled with $J_1t$. On the other hand, with finite-range hopping changes to the spin-ordering are fundamentally slower than $1/J_1$ or $1/J_2$, consistent with the fact that our entanglement dynamics at long times does not collapse upon rescaling with either hopping rate (we show the entanglement dynamics do not collapse with a rescaling of time for $\alpha = 0.5$ and $6$ in Appendix \ref{sec:app2}). Lastly, we emphasize that our observation of two-stage dynamics and the slow relaxation of the system at long times is not specific to our initial state. We have observed a fast period of transient dynamics followed by slow growth of entanglement across multiple decades for other generic initial states when $J_1/J_2 \ll 1$, such as, e.g., the CDW state used in Figs.~\ref{fig:longrange} and \ref{fig:Ssc}. 

We comment briefly on the relation of our observations to prethermalization, which is characterized by a separation of timescales involving quick equilibration to a long-lived prethermal state followed by eventual relaxation to true thermal equilibrium. Many prior studies of prethermal behaviour have focused on systems featuring an integrable limit, i.e., possessing an extensive set of conserved quantities, which are perturbed by an integrability-breaking term \cite{Berges2004prethermalization,Ueda2018prethermalization,Ueda2020prethermalization}. At short times the dynamics is constrained within the reduced Hilbert space of the integrable part of the Hamiltonian and local observables equilibrate to the predictions of a generalized Gibbs ensemble determined by the associated conserved quantities. Nevertheless, over extended time scales, the system thermalizes completely by leaking into the larger Hilbert space of the complete model. This process occurs with a relaxation timescale proportional to $1/g^2$, in accordance with Fermi's golden rule, where $g$ represents the characteristic strength of the non-integrable perturbation. Although in the limit $J_1 = 0$ our model is non-integrable, our understanding of the two-stage relaxation is clearly analogous to prethermalization. In fact, recent work has demonstrated that non-integrable systems featuring a set of conserved quantities and subject to a weak perturbation that removes at least one of these quantities can still feature prethermal behaviour \cite{Marcos2018prethermalization,Marcos2019prethermalization,Marcos2021prethermalization}. For our system in the limit $J_1 = 0$, the initial configuration of the immobile species forms a set of conserved quantities, which is broken for slow hopping $J_1 \neq 0$. However, as previously mentioned, for the relatively modest system sizes that we probe neither the entanglement entropy or local observables such as the central charge density $\bar{n}$ exhibit relaxation with the characteristic scaling $J_1^2 t$.

\section{Quench dynamics of $\hat{H}^{\prime}_{\mathrm{BH}}$}\label{sec:modelbd}
The prior study of $\hat{H}_{\mathrm{BH}}$ in Sec.~\ref{sec:modelud} has demonstrated that strongly imbalanced hopping rates for the two components can lead to dynamics featuring multiple stages with distinct timescales. In this section, we show that the alternative mapping of the Rydberg Hamiltonian with hardcore bosons that realizes $\hat{H}^{\prime}_{\mathrm{BH}}$ [Eq.~(\ref{eq:bd})] can provide a complementary example of multi-stage dynamics, even when the hopping rates $J_1$ and $J_2$ are almost equal. 

Notably for $\hat{H}^{\prime}_{\mathrm{BH}}$, breaking the symmetry between the dipolar interactions, $J_1 \neq J_2$, now explicitly breaks the separate conservation of $N_{\uparrow}$ and $N_{\downarrow}$, although the total boson number remains fixed. This change is driven by the spin-flip hopping process in $\hat{H}^{\prime}_{\mathrm{BH}}$ that occurs with a rate $J^f_{ij} = J_f/r_{ij}^3$ between sites $i$ and $j$ where $J_f = (J_1 - J_2)/2$. A related consequence of $J_1 \neq J_2$ is that the evolution of the spin-ordering is no longer solely driven by beyond-nearest-neighbour contributions to the hopping of each component, as the spin-flip term in $\hat{H}^{\prime}_{\mathrm{BH}}$ can change the spin-ordering even via a nearest-neighbour hop onto an empty site. Throughout this section we will assume $J_1, J_2 \geq 0$ and thus $J_f/J_h \in [-1,1]$ for simplicity.

\begin{figure}[bt]
\begin{center}
\includegraphics[width=1\columnwidth]{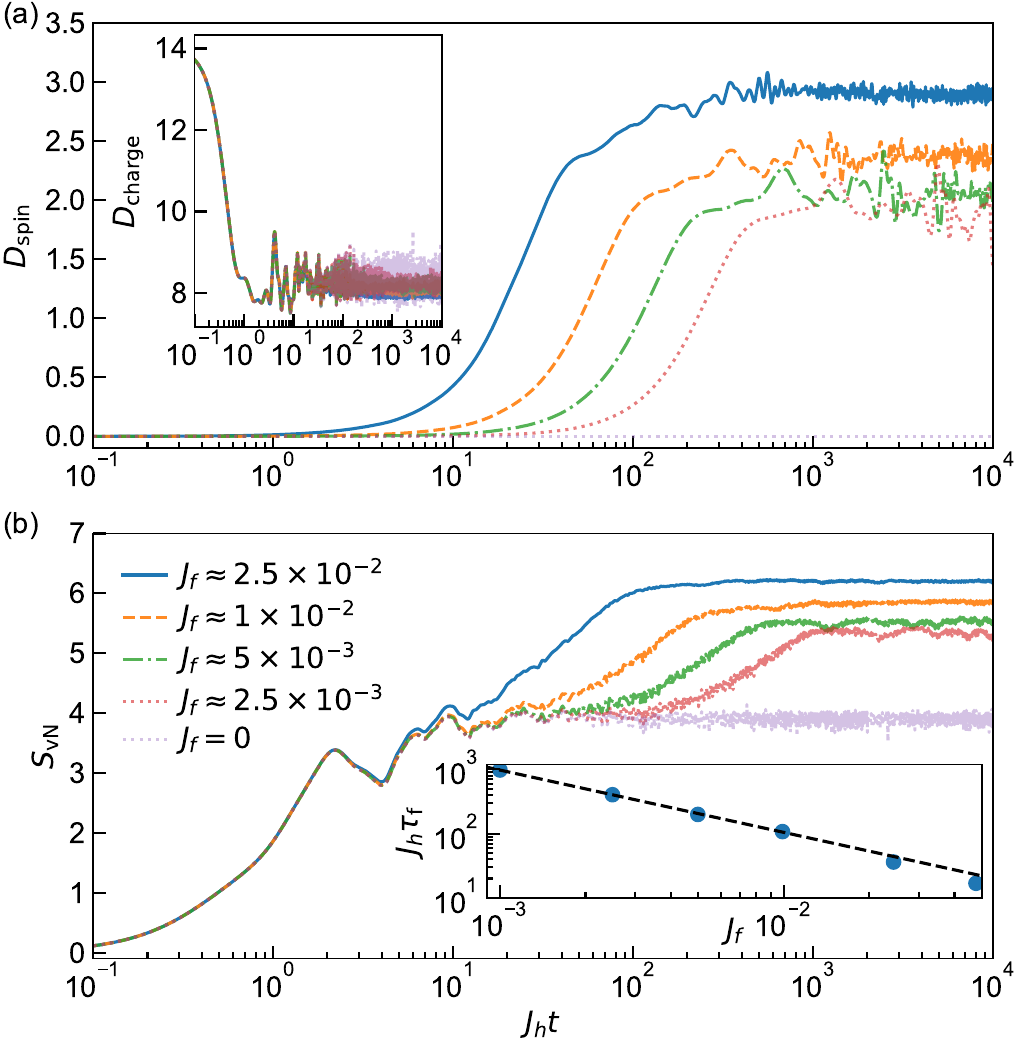}
\caption{Relaxation dynamics for weak spin-flip hopping $J_f \ll J_h = 1$ and initial state $\ket{\uparrow0\uparrow0\cdots}$ with periodic boundary conditions and $\mathcal{N} = 14$. Values of $J_f$ are indicated in legend of panel (b). (a) Relaxation of $D_{\rm spin}$ (main panel) and $D_{\rm charge}$ (inset) as a function of rescaled time $J_h t$. (b) Spatial entanglement entropy $S_{\rm vN}$ (main panel) as a function of rescaled time $J_h t$. Inset: Characteristic relaxation time $\tau_f$ calculated from $S_{\rm vN}$ (see main text). The dashed line is obtained by fitting $\tau_f \sim (J_f/J_h)^{-a}$ to the data for $J_f\leq10^{-2}$ for which $a = 0.97$. 
} 
\label{fig:bd_smalldiff}
\end{center}
\end{figure}

Figure \ref{fig:bd_smalldiff} presents numerical simulations of the dynamics generated by $\hat{H}^{\prime}_{\mathrm{BH}}$ in the limit of $J_f \ll J_h$ (i.e., $J_1 \approx J_2$). To emphasize the role of spin-flip hopping we use an initial state $|\uparrow0\uparrow0\cdots\rangle$ with periodic boundary conditions. All simulations in this section use $\alpha=3$. Panel (a) shows the dynamics of spin and charge observables $D_{\rm spin}$ (main panel) and $D_{\rm charge}$ (inset), respectively. The relaxation of the spin degree of freedom occurs on a characteristic timescale $t \sim 1/J_f$ and $D_{\mathrm{spin}}$ equilibrates at long times to a value consistent with the diagonal ensemble. The initial relaxation of the charge degree of freedom, which occurs on the much faster timescale $1/J_h$, is independent of the spin degree of freedom for for $J_h t < J_h/J_f$, as evidenced by the indistinguishability of $J_f \neq 0$ data with a calculation that explicitly uses $J_f = 0$ (magenta dashed line). At long times $D_{\mathrm{charge}}$ oscillates around a value (consistent with a diagonal ensemble calculation) that weakly depends on $J_f$. Collectively, our results indicate that the relaxation of each degree of freedom effectively proceeds independently and is driven separately by the spin-independent ($\propto J_h$) and spin-flip ($\propto J_f$) hopping terms of $\hat{H}^{\prime}_{\mathrm{BH}}$, respectively.

Figure \ref{fig:bd_smalldiff}(b) shows corresponding data for the spatial entanglement entropy $S_{\rm vN}$.  Consistent with the prior discussion of panel (a), we observe that the initial dynamics for $J_f \neq 0$ follows the expectations for the single component model ($J_f = 0$), with entanglement building up quickly on a timescale $\sim 1/J_h$. A second period of slower growth of entanglement occurs for $J_h t \gtrsim J_h/J_f$ and, consistent with the relaxation of $D_{\mathrm{spin}}$, the growth is dominated entirely by the spin-flip hopping. This latter conclusion is supported by extracting a characteristic relaxation time $\tau_f$ defined as that at which $S_{\rm vn}$ reaches $[S_{\rm vN, f}-S_{\rm vN, f}(J_f=0)]/2$, i.e., halfway between the long-time values $S_{\rm vN, f}$ and $S_{\rm vN, f}(J_f=0)$ with $J_f \neq 0$ and $J_f = 0$, respectively. Here, we numerically estimate the long-time value as the time-average of $S_{\mathrm{vN}}$ over the interval $10^4 \leq J_h t \leq 2 \times 10^4$. The inset of panel (b) shows $\tau_f$ as a function of $J_f$. We fit the results to a power-law $\tau_f \sim (J_f/J_h)^{-a}$ and obtain $a = 0.97$ indicating an approximately inversely proportional relationship, which confirms that spin-flip hopping drives the growth of entanglement.

\begin{figure}[bt]
\begin{center}
\includegraphics[width=1\columnwidth]{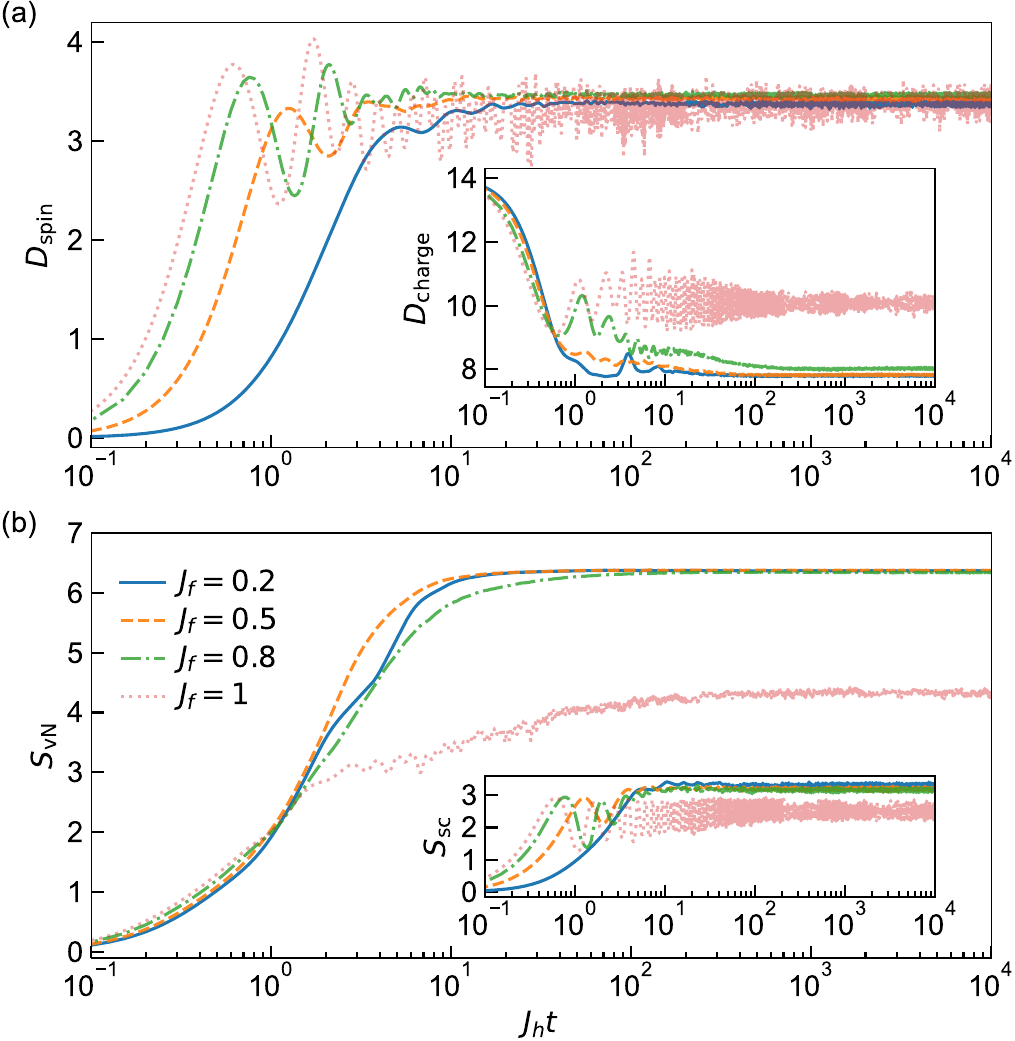}
\caption{Relaxation dynamics for $J_f \approx J_h = 1$  and initial state $\ket{\uparrow0\uparrow0\cdots}$ with periodic boundary conditions and $\mathcal{N} = 14$. Values of $J_f$ are indicated in legend of panel (b). (a) Relaxation of $D_{\rm spin}$ (main panel) and $D_{\rm charge}$ (inset) as a function of rescaled time $J_h t$. (b) Spatial entanglement entropy $S_{\rm vN}$ (main panel) and spin-charge entanglement entropy $S_{\rm sc}$ (inset) as a function of rescaled time $J_h t$.
} 
\label{fig:bd_largediff}
\end{center}
\end{figure}

We also probe the limit where $J_1 \gg J_2$, corresponding to $J_f \sim J_h$, in Fig.~\ref{fig:bd_largediff}. Panel (a) shows the evolution of $D_{\rm spin}$ (main panel) and $D_{\rm charge}$ (inset) for increasing ratios of $J_f/J_h = 0.2, 0.5, 0.8$ and $1$. We observe that the delineation of the characteristic spin and charge timescales is diminished as $J_f/J_h$ increases, consistent with expectations that the propagation of the bosons through the lattice is almost equally probable to be accompanied with or without a spin-flip. The dynamics of the entanglement entropies $S_{\rm vN}$ [Fig.~\ref{fig:bd_largediff} (b) main panel] and $S_{\rm sc}$ [Fig.~\ref{fig:bd_largediff} (b) inset] also reflect this. The former no longer shows distinct two-step evolution,  while the latter shows that spin-charge entanglement builds up increasingly quickly commensurate with $J_f/J_h$. We highlight that the long time behaviour of $J_f/J_h=1$ is quite different in all panels of Fig.~\ref{fig:bd_largediff}. In that case, large oscillations are observed in $D_{\rm spin}$, $D_{\rm charge}$, and $S_{\rm sc}$ about well-defined mean values that significantly deviate from the other choices of $J_f/J_h$ that we show. This distinct behaviour can be understood by instead viewing the dynamics through the lens of the alternative Hamiltonian $\hat{H}_{\mathrm{BH}}$. In that frame, one component becomes completely immobile ($J_2 = 0$) and the configuration of these immobile bosons introduces a constraint on the hopping dynamics of the mobile component and the accessible Hilbert space.

\section{Experimental considerations}\label{sec:experiment}
The physics discussed in the prior sections is within reach of state-of-the-art quantum simulators based on tweezer arrays of Rydberg atoms. For example, the preparation of arbitrary product states with site-resolved structure (such as Neel spin ordering or density waves) can be achieved by exploiting light shifts of the initially trapped ground state or Rydberg levels \cite{BrowaeysSymmetryBreaking}, combined with independent coherent driving of each Rydberg-Rydberg transition with tunable microwave fields~\cite{kaden2023rydbergsynthetic, sundar2019synthetic}. This coherent control of each atom, combined with high fidelity site-resolved imaging \cite{Ebadi2021,Madjarov2020, Evered2023,bornet2024enhancing, manetsch2024tweezer}, can enable the extraction of many-body correlations and the construction of, e.g., number entropies. We note that the extraction of charge observables only requires the measurement of population in a single Rydberg level, which can be achieved by standard detection techniques that transfer population of the targeted Rydberg level back to the ground-state for recapture in the tweezers and subsequent imaging~\cite{JauNP}. For spin observables, one can in principle measure the populations in all three Rydberg levels by selective transfer and shelving of population of each Rydberg state into different ground-state magnetic sublevels~\cite{LeeJCladder}.

Observing the relaxation dynamics over long timescales requires the identification of Rydberg levels with sufficiently strong dipolar exchange interactions compared to relevant technical and fundamental limitations on the total simulation time. One such constraint is residual thermal motion of the atoms that limits recapture of the atoms by the tweezers for subsequent imaging (the tweezers are turned off during the quench dynamics to avoid perturbation of the Rydberg states). For the micro-Kelvin temperatures routinely achieved in state-of-the-art tweezer experiments this leads to a first limit of about $t \lesssim 20~\mu$s. A second constraint is the finite lifetime of the chosen Rydberg states. While the effects of this can in principle be included in numerical simulations comparing to a specific experimental apparatus, this is beyond the scope of the current work. Instead, we highlight that any effects can be suppressed by a combination of state-resolved single-shot detection (see above) and post-selection. In particular, state-resolved detection of all three Rydberg states in a single experimental shot enables discrimination of cases where an atom decays out of the targeted Rydberg levels during the quench dynamics. This approach is feasible as Rydberg quantum simulators typically have relatively fast duty cycles and the fraction of experimental shots where no loss occurs can be sufficiently large for the moderate sized arrays discussed in this work. The latter is approximately quantified by the survival probability for $\mathcal{N}$ atoms, $P_{\mathcal{N}} = e^{-\mathcal{N}t/\tau}$, where $t$ is the quench duration and $\tau$ is a characteristic lifetime of the relevant Rydberg states.

As an illustrative example, Table~\ref{table:param} shows relevant parameters for candidate Rydberg levels in cesium. The states and related parameters are chosen by requiring that timescales of the relevant physics (motivated by Figs.~\ref{fig:unequalhopping} and \ref{fig:bd_smalldiff}) can be reached with a survival probability of at least $P_{\mathcal{N}} \geq 0.1$ with $\mathcal{N} = 10-20$, as well as other factors such as feasible tweezer separation, reasonable simulation time (i.e., sufficient recapture probability as discussed above) and suppressed additional short-range van der Waals interactions $J_{\mathrm{vdW}}/r_{ij}^6$ neglected in our model \cite{morgado2021quantum}. Moreover, additional off-resonant dipolar interactions to other Rydberg state magnetic sublevels can be feasibly suppressed with an appropriately strong quantization field \cite{LeeJCladder,Evered2023}. In general, a far greater diversity of possible level configurations can be explored, including by consideration of other atomic species, with an eye to tuning the precise values of $J_{1,2}$ (and associated parameters) or optimizing for other regimes (i.e., shorter timescales such as those probed in Fig.~\ref{fig:longrange}).

\begin{table*}[tb!]
\begin{ruledtabular}
\begin{tabular}{ccccccc}
Configuration & State & Lifetime~$\left(\rm \mu s \right)$ & Level   & $C_{3}$/h$~\left(\rm MHz \cdot \mu m^3 \right)$ & Spacing~$\left(\rm \mu m \right)$ &  $J_i/2\pi~(\rm MHz)$  \\

\hline
& $|1\rangle$ & 51 & 51F$_{5/2, mj=1/2}$ &  379, ($|1\rangle \leftrightarrow |2\rangle$) & &   $J_1/{2\pi}$ = 0.74  \\
$J_1/J_2 =0.082$ & $|2\rangle$ & 55&  53D$_{3/2, mj=3/2}$ &  4625, $\left(|2\rangle \leftrightarrow |3\rangle\right)$ & 8 & $J_2/{2\pi}$ =  9.03  \\
& $|3\rangle$ & 121 &  54P$_{1/2, mj=1/2}$& &     &   \\
\hline
& $|1\rangle$ & 51  & 51F$_{7/2, mj=3/2}$ &  1973, ($|1\rangle \leftrightarrow |2\rangle$) & &  $J_1/{2\pi}$ = 2.71  \\
$J_1/J_2 =0.877$ & $|2\rangle$ & 55 & 53D$_{5/2, mj=3/2}$ &   2246,  ($|2\rangle \leftrightarrow |3\rangle$) & 9 &  $J_2/{2\pi}$ = 3.08 \\
& $|3\rangle$ & 118 & 54P$_{3/2, mj=3/2}$&    &  &    \\
\hline
& $|1\rangle$ & 84 &  
 62D$_{5/2, mj=1/2}$&  3244.8, ($|1\rangle \leftrightarrow |2\rangle$) & &  $J_1/{2\pi}$ = 0.7922  \\
$J_1/J_2 =1.004$ & $|2\rangle$ & 169 & 63P$_{3/2, mj=1/2}$ &   3232.3,  ($|2\rangle \leftrightarrow |3\rangle$) & 16 &  $J_2/{2\pi}$ = 0.7891 \\
& $|3\rangle$ & 112& 64S$_{1/2, mj=1/2}$&    &  &    \\
\end{tabular}
\end{ruledtabular}
\caption{\label{table:param}
Summary of candidate Rydberg states to achieve target values of $J_1/J_2$ \cite{ARC_library}. For the case of $J_1/J_2 = 0.082$ the total survival probability for the atom array is in the range of about $0.5 < P_{\mathcal{N}} < 0.7$ for $10 \leq \mathcal{N} \leq 20$ at $J_2t=100$. Similarly, for $J_1/J_2 = 0.877$ we have $0.13 < P_{\mathcal{N}} < 0.36$ at $J_2t=100$, and for $J_1/J_2 = 1.004$ we have $0.24 < P_{\mathcal{N}} < 0.49$ at $J_2t=30$, both using the same range of array sizes.
}
\end{table*}

\section{Summary and outlook}\label{sec:summary}

Our work demonstrates the opportunities for quantum simulation and studies of non-equilibrium dynamics with multilevel Rydberg atoms. In our example using three internal states, we showed the tunability of the platform can enable controllable studies of a two-component Bose-Hubbard model with tunable, power-law hopping and hardcore interactions, which features rich relaxation dynamics that depend on the interplay of these ingredients. Beyond prior work studying dynamical constraints, localization and thermalization in this and similar models \cite{Gadway2011glassy,Sarkar_2020,Garrahan2018dynamicalconstraints,Muller2015mbl,Yao2016quasimbl,Sirker2019quasimbl,Folling2022massimbalanced}, we also highlight potential connections to recent studies of glassy dynamics in ensembles of Rydberg qubits featuring positional disorder \cite{signoles2021glassy}. For instance, preparing an initial product state where one component is immobile (i.e., $J_1/J_2 = 0$) and is randomly scattered to populate a variable fraction of the chain while the remaining sites are in a superposition of empty and occupied by the mobile component, our model directly maps to typically studied configurations. An interesting direction would be to perturb away from this scenario by allowing the previously immobile component to slowly move ($J_1/J_2 \ll 1$). Note that the predictions we discuss in this work can also be realized in other systems featuring multilevel internal structure and exchange interactions, such as polar molecules \cite{sundar2018synthetic,sundar2019synthetic,Cornish2024Quantum}.

The versatility and breadth of the Rydberg platform also suggests a number of interesting extensions for future work. For example, while we have already demonstrated that the choice of basis states in the three-level system can be used to generate distinct spin-flip hopping, working at small tweezer spacings to purposely strengthen the van der Waals interactions between Rydberg atoms can be used to engineer additional $1/r^6$ density-density interactions in the Bose-Hubbard model that can compete with the $1/r^3$ hopping. Microwave driving of the different Rydberg transitions can also enable studies of Floquet-engineered Hamiltonians in the spirit of recent work with Rydberg qubits and polar molecules \cite{geier2021ultracold,weidemuller2022floquet,christakis2023probing,Nishad2023floquet, Homeier2023Antiferromagnetic}. Lastly, all of these tools can similarly be applied to study non-equilibrium dynamics in higher-dimensional arrays and ladders of more than three Rydberg states, although for the latter a description in terms of hardcore bosons is no longer readily applicable.

\section{Acknowledgment}
This material is based upon work supported by the Air Force Office of Scientific Research under award numbers FA9550-22-1-0335 and FA9550-20-1-0123. Y.~Z also thanks the support from the Dodge Family Postdoctoral Fellowship at the University of Oklahoma.

\appendix

\section{Long-time relaxation of spin and charge observables} \label{sec:app1}
\begin{figure}[bt!]
\begin{center}
\includegraphics[width=1\columnwidth]{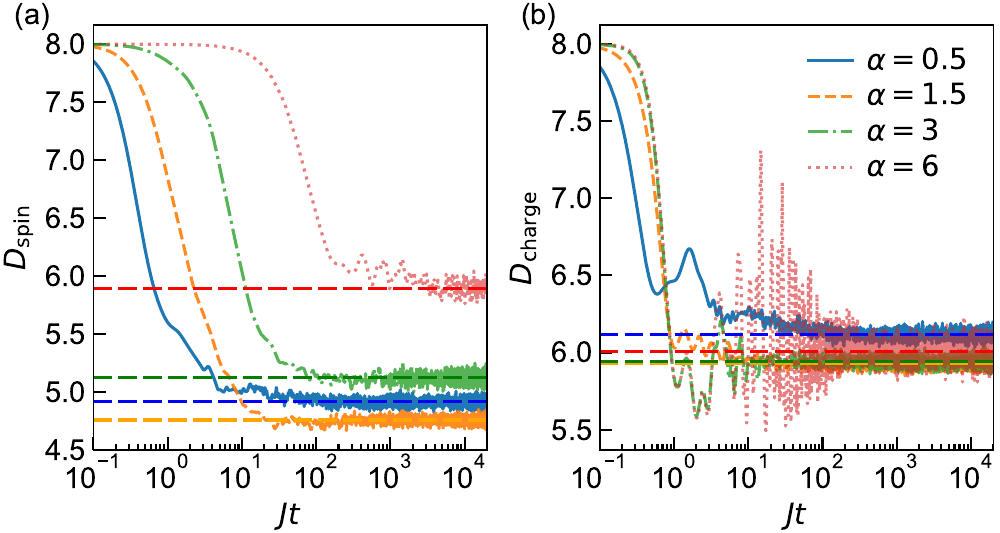}
\caption{Long-time quench dynamics for $\hat H_{\rm BH}$ [Eq.(\ref{eq:ud})] with $J_1 = J_2 = J$ and  starting from the charge density wave initial state $\ket{\varphi_{\mathrm{cdw}}} = \ket{\uparrow\downarrow0\uparrow\downarrow0\cdots}$ with periodic boundary conditions and $\mathcal{N} = 12$. (a) and (b) Evolution of $D_{\rm spin}$ and $D_{\rm charge}$ for a range of hopping exponents $\alpha$ [see legend in panel (b)]. Horizontal dashed lines [colors according to the legend in panel (b)] indicate the predictions for the associated diagonal ensemble.} 
\label{fig:app1}
\end{center}
\end{figure}

As the model $\hat H_{\rm BH}$ [Eq.~(\ref{eq:ud})] is non-integrable, it is expected that for a sufficiently large system the long-time values of local observables should be consistent with the predictions of an infinite temperature canonical ensemble (as our initial state has zero mean energy). In particular, the canonical ensemble predicts equilibrium values for number of charge and spin domains, $D_{\rm charge}=\frac{(N_{\uparrow}+N_{\downarrow})^2}{{\cal N}-1}$ and $D_{\rm spin}=\frac{(N_{\uparrow}+N_{\downarrow})^2}{2(N_{\uparrow}+N_{\downarrow}-1)}$, respectively. However, due to our limited system size (${\cal N} = 15$ in Fig.~\ref{fig:longrange}), we expect deviations due to finite size effects. Instead, our system should relax to the diagonal ensemble prediction \cite{dalessio16review} (which is expected to coincide with the canonical ensemble as ${\cal N} \to \infty$ if the system thermalizes, see below).  Exact calculation of the diagonal ensemble is computationally expensive for ${\cal N}\geq15$. Thus, to confirm that at long times our results match the predictions of the diagonal ensemble, in Fig.~\ref{fig:app1} we plot the quench dynamics for $D_{\rm spin}$ and $D_{\rm charge}$ from the same initial state $\ket{\varphi_{\mathrm{cdw}}} = \ket{\uparrow\downarrow0\uparrow\downarrow0\cdots}$ as in main text, but for the smaller system size ${\cal N}=12$. Examination of the data in Fig.~\ref{fig:app1} shows agreement between the predictions of diagonal ensemble (horizontal dashed lines) and the saturated values of $D_{\rm charge}$ and $D_{\rm spin}$ at sufficiently long times (set by the hopping range).

\begin{figure}[bt!]
\begin{center}
\includegraphics[width=1\columnwidth]{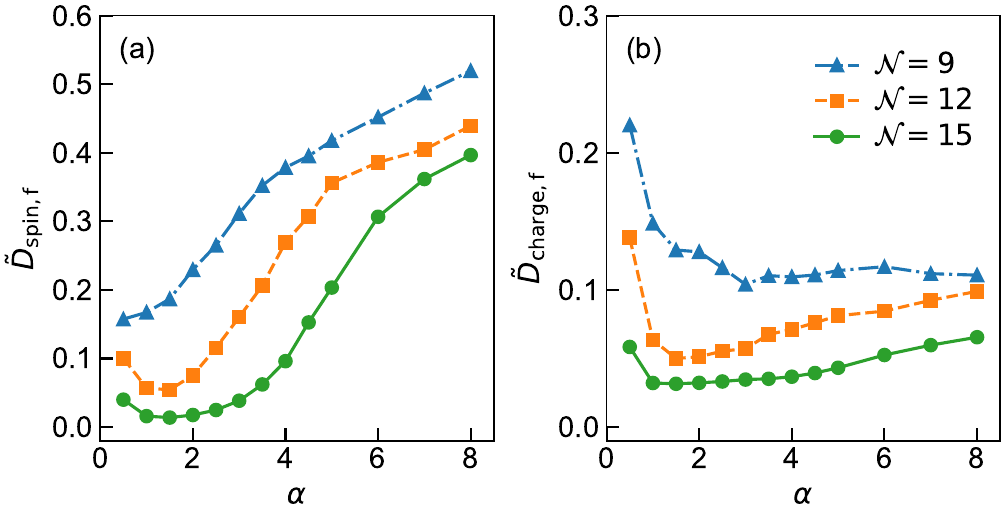}
\caption{(a) $\tilde D_{\rm spin,f}$ [see Eq.~(\ref{eq:dfscale})] and (b) $\tilde D_{\rm charge,f}$ as a function of hopping exponent $\alpha$. Initial state is $\ket{\varphi_{\mathrm{cdw}}}$ as per Fig.~\ref{fig:app1} but we show data for system sizes $\cal N$=9, $12$, and $15$ [see legend in panel (b)].
} 
\label{fig:app2}
\end{center}
\end{figure}

To systematically assess the role of finite size effects in the long-time equilibration of our system, we also compare the predictions of the diagonal ensemble and the infinite temperature canonical ensemble as a function of $\cal N$. We study the same quench dynamics from the density wave initial state and show results for system sizes ${\cal N} =9$, $12$, and $15$ in Fig.~\ref{fig:app2}. To appropriately compare the ensembles, we define the rescaled quantity,
\begin{equation}\label{eq:dfscale}
    \tilde D_{\rm spin, f}=\frac{D_{\rm spin, f}-D_{\rm spin}(\infty)}{D_{\rm spin}(0)-D_{\rm spin}(\infty)}\,,
\end{equation}
where $D_{\rm spin}(0)$ is the initial ($t=0$) value of $D_{\rm spin}$ and $D_{\rm spin}(\infty)$ is the prediction of the infinite temperature canonical ensemble. We obtain the relaxed value $D_{\rm spin, f}$ via the long-time average $D_{\rm spin, f} = \frac{1}{T}\int_{T}^{2T} D_{\rm spin}(t) dt$ with $T = 10^4 J^{-1}$, consistent with our analysis in Fig.~\ref{fig:longrange} of the main text. We similarly define the rescaled charge observable $\tilde D_{\rm charge, f}$. The definition of $\tilde D_{\rm spin, f}$ ($\tilde D_{\rm charge, f}$) is chosen such that it approaches unity if the number of spin (charge) domains is conserved (as is the case for $\alpha\to\infty)$ and vanishes if the diagonal and canonical ensemble predictions  coincide (i.e., as expected for ${\cal N}\to\infty$). For both spin and charge dynamics, $D_{\rm spin, f}$ and $D_{\rm charge, f}$ become closer to the canonical ensemble prediction as $\cal N$ increases. We note that as the hopping range is decreased, we find larger discrepancies between $D_{\rm spin,f}$ and $D_{\rm spin} (\infty)$ at a given $\cal N$, consistent with the fact we expect a discontinuity between nearest-neighbour hopping and finite $\alpha$ in the limit of large system size. On the other hand, the predictions for the number of charge domains does not significantly depend on $\alpha$.

\begin{figure}[bt]
\begin{center}
\includegraphics[width=1\columnwidth]{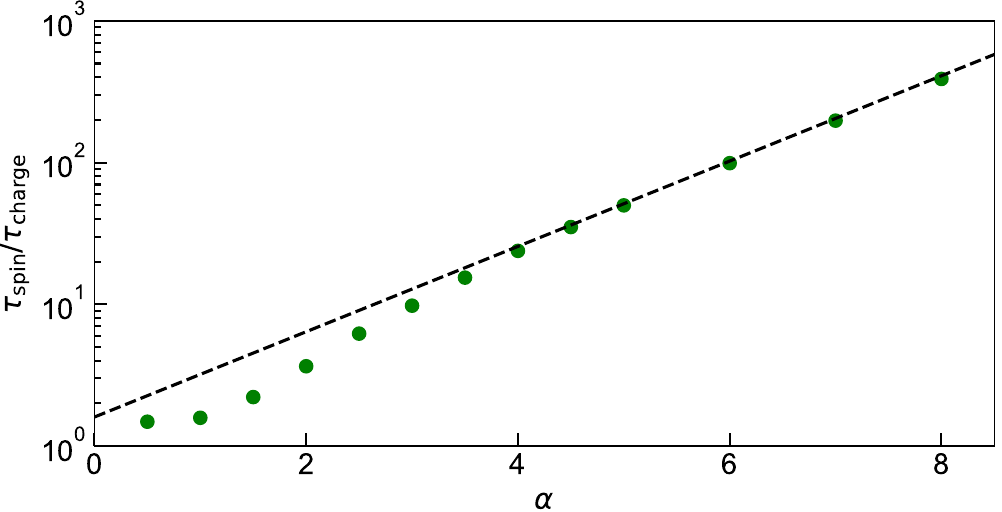}
\caption{Ratio of relaxation timescales $\tau_{\rm spin}/\tau_{\rm charge}$ as a function of hopping exponent $\alpha$. The timescales are obtained from the data of Fig.~\ref{fig:longrange}(d) and (e) in the main text. Dashed line plots $\tau_{\rm spin}/\tau_{\rm charge}\sim 2^{\alpha}$ to guide the eye.} 
\label{fig:app3}
\end{center}
\end{figure}

We conclude this appendix with a brief discussion of the spin and charge timescales, $\tau_{\rm spin}$ and $\tau_{\rm charge}$, that are plotted in Fig.~\ref{fig:longrange} of the main text. Further insight into their behaviour with hopping range is obtained by plotting the ratio of $\tau_{\rm spin}/\tau_{\rm charge}$ in Fig.~\ref{fig:app3}. Deep inside the short-range regime where $\alpha>>2$, the ratio follows an exponential form, $\tau_{\rm spin}/\tau_{\rm charge}\sim 2^{\alpha}$ (dashed line). Gradual deviation from this exponential form is observed as the boundary between short and long-range regimes ($\alpha \sim 2$) is approached. For very long-range hopping $\alpha\leq1$ we find $\tau_{\rm spin}/\tau_{\rm charge}\sim 1$ as the spin and charge scales are no longer well separated by nearest and next-nearest neighbour hopping rates.

\section{Influence of hopping range on observation of slow relaxation} \label{sec:app2}

\begin{figure}[bt]
\begin{center}
\includegraphics[width=1\columnwidth]{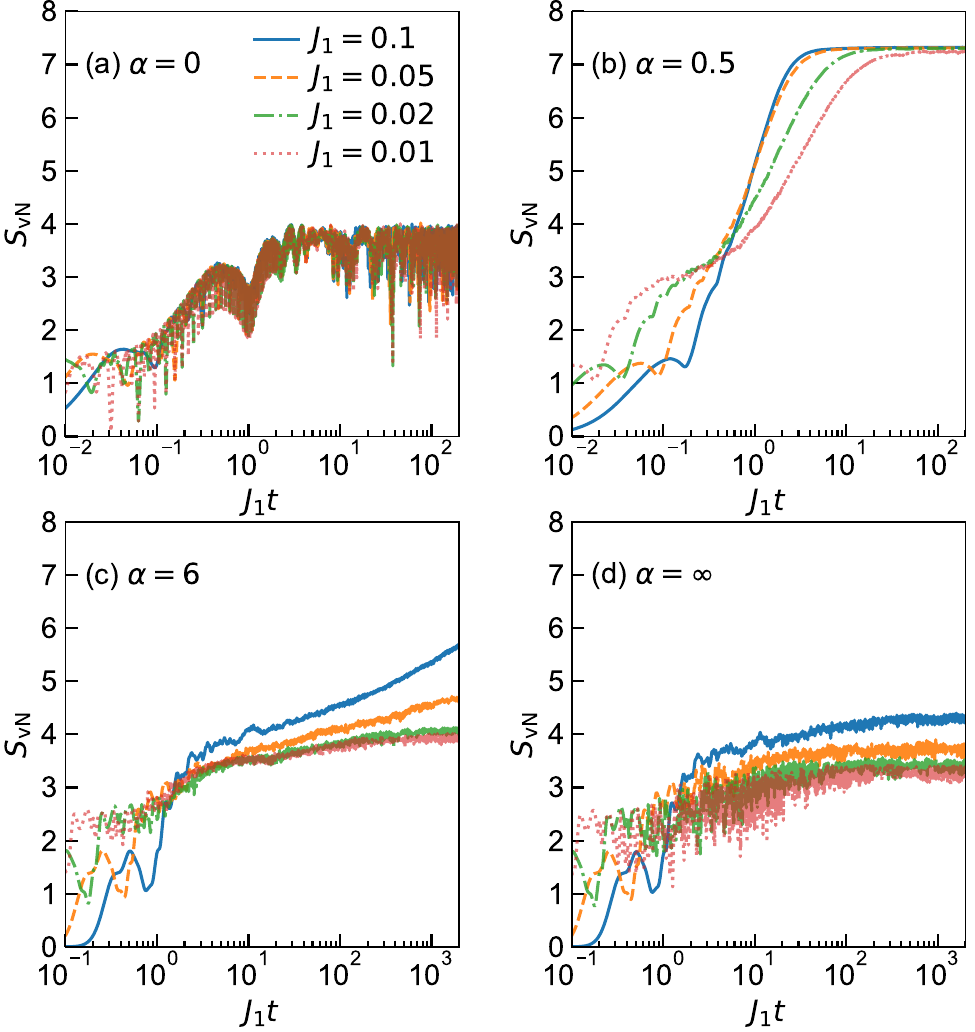}
\caption{Growth of entanglement $S_{\rm vN}$ in the regime of strongly imbalanced hopping for exponents (a) $\alpha=0$, (b) $\alpha=0.5$, (c) $\alpha=6$, and (d) $\alpha = \infty$ (nearest-neighbor hopping). Values of $J_1$ are indicated in legend of panel (a) and $J_2 = 1$ is fixed for all data. Note that we plot all data as a function of rescaled time $J_1 t$.} 
\label{fig:smallJalpha}
\end{center}
\end{figure}

In Sec.~\ref{sec:hopping} of the main text, we observe that when the hopping amplitudes of the two components are very different, i.e., $J_1 << J_2$, the relaxation can be decomposed into a period of rapid transient dynamics on timescales $1/J_2$ and a second, slower stage involving the transport of both components. The slow dynamics is driven by an interplay of the disparate hopping rates, hard-core interactions and power-law tail of the interactions. To demonstrate the latter, here we study the relaxation dynamics for a range of different hopping exponents $\alpha$. The different panels of Fig.~\ref{fig:smallJalpha} show results for the entanglement entropy $S_{\rm vN}$ as a function of rescaled time $J_1t$, analogous to the inset of main text Fig.~\ref{fig:slowJ1}(a), for $\alpha=0$, 0.5, 6, and $\infty$. In the case of all-to-all hopping ($\alpha = 0$) the geometry of the 1D chain and the initial spin configuration become effectively irrelevant, evidenced by the fact that the dynamics at long times ($>1/J_2$) collapses for different hopping rates when time is rescaled as $J_1t$. On the other hand, for any value of $\alpha\neq0$, we observe that the dynamics does not collapse upon rescaling with $J_1$. Moreover, the slowdown of the relaxation becomes increasingly pronounced as $\alpha$ is increased.

\section{Role of sign of hopping in $\hat H'_{\rm BH}$} \label{sec:app3}
In the main text Sec.~\ref{sec:modelbd}, we studied the role of the hopping imbalance in the dynamics of $\hat H^{\prime}_{\rm BH}$ but fixed the sign $J_1, J_2 > 0$ such that $J_{\rm f}/J_{\rm h}=(J_1-J_2)/(J_1+J_2) \in (-1, 1)$. However, the dipolar interactions $J_1$ and $J_2$ do not have a fixed relative sign and thus a broader range of $J_{\mathrm{h}}$ and $J_{\mathrm{f}}$ can in principle be accessed. For example, when $J_1 > 0$ and $J_2 < 0$ the denominator $|J_1+J_2|$ is smaller than $|J_1-J_2|$, such that $|J_{\rm f}/J_{\rm h}|\in(1,\infty)$. By controlling the relative sign of the dipolar interactions, we can thus experimentally access any value of $J_{\rm f}/J_{\rm h}$, including $J_{\rm h}=0$ ($J_1=-J_2$). However, we choose to not explore values outside the range $J_{\rm f}/J_{\rm h} \in (-1, 1)$ in this work. 

\bibliographystyle{apsrev4-1}
\bibliography{Reference}

\end{document}